\documentclass[aps,preprint]{revtex4}

\usepackage{epsfig,amsmath}
\usepackage{subfigure}
\usepackage{graphicx}
\usepackage{dcolumn}
\usepackage{stmaryrd}
\usepackage{mathrsfs}
\usepackage{pifont}
\usepackage{amsthm}
\usepackage{amssymb}
\usepackage{bm}
\usepackage{latexsym}
\usepackage[colorlinks=true,linkcolor=blue,citecolor=blue]{hyperref}
\usepackage{color}

\theoremstyle{plain}

\newcommand{\la}{\langle}
\newcommand{\ra}{\rangle}

\newcommand{\ti}{\tilde}
\newcommand{\ga}{\gamma}

\newcommand{\da}{\dagger}

\newcommand{\al}{\alpha}
\newcommand{\si}{\sigma}

\newcommand{\om}{\omega}

\newcommand{\de}{\delta}

\newcommand{\non}{\nonumber}
\newcommand{\pa}{\partial}

\def\pra#1{{ Phys.\ Rev. A\/} {\bf#1}}
\def\prb#1{{ Phys.\ Rev. B\/} {\bf#1}}

\def\prl#1{{ Phys.\ Rev.\ Lett.} {\bf#1}}

\def\pla#1{{ Phys.\ Lett. A\/} {\bf#1}}

\def\nat#1{{ Nature} {\bf#1}}

\def\njp#1{{ New. J. \ Phys.} {\bf#1}}

\begin{document}

\title{Creating Quantum Cluster States embedded in ancillary baths}

\author{Jun Jing$^{1,2,3}$, Mark S. Byrd$^{4}$, and Lian-Ao Wu$^{2}$\footnote{Corresponding author: lianao.wu@ehu.es}}

\affiliation{$^{1}$Department of Physics, Zhejiang University, Hangzhou 310027, Zhejiang, China \\ $^{2}$Department of Theoretical Physics and History of Science, The Basque Country University (EHU/UPV), PO Box 644, 48080 Bilbao, and Ikerbasque, Basque Foundation for Science, 48011 Bilbao, Spain \\ $^{3}$Institute of Atomic and Molecular Physics, Jilin University, Changchun 130012, Jilin, China \\ $^{4}$ Physics Department, Southern Illinois University, Carbondale, IL, 62901-4401, USA}

\date{\today}

\begin{abstract}
We propose a systematic and explicit method for the inverse engineering of the dynamics of an open quantum systems with no auxiliary Hamiltonian nor the prerequisite of adiabatic passage. In particular, we exploit the Lindblad dissipators in order to create a resource state or subspace of interest in the presence of decoherence. In a conceptual shift, the Lindblad dissipators, including multiple interactions that are central to determine the steady state in the long-time limit for an open quantum system, can be guided to produce a useful practical resource to achieve an {\em arbitrary} target state or subspace. More importantly, with the help of gate and circuit-based quantum control, we provide an explicit, programmable, and polynomially efficient control sequence to create a cluster state or graph state useful for one-way quantum computing.
\end{abstract}


\maketitle

A prominent question in both fundamental and applied quantum information science is whether a quantum state can survive for long enough time scales in the presence of a deleterious environment to enable the implementation of a protocol or algorithm. The need to tailor and control the dynamics of open quantum systems has led to the discovery of a host of techniques to palliate and avoid the effects of decoherence~\cite{Breuer,Lidar15}, such as dynamical decoupling~\cite{dd1,dd2}, quantum error correction~\cite{Lidar/Brun,Gaitan}, decoherence-free subspaces (DFS)~\cite{DFS1,DFS2}, noiseless subsystems~\cite{nl} and holonomic quantum computation~\cite{holo}. More recently, it has been appreciated that quantum dissipative processes might be able to be controlled and guided to produce resources for quantum information processing tasks~\cite{DP1,DP2,DP3,DP4,DP5}. Therefore, knowledge of the energy structure and the dissipation manner of the system can be exploited to allow the creation of an on-demand subspace or state~\cite{Mark15}. It therefore enables the inverse engineering~\cite{inv1,inv2} in the regime of open-system quantum dynamics by control over system variables, avoiding the requirement of an auxiliary Hamiltonian~\cite{DR03,DR05,Berry} and/or the adiabatic condition~\cite{adiab1,adiab2}.

Most of the existing experiments on quantum control are designed to implement sequences of highly structured interactions between selected qubits, thereby following the quantum circuit model of a universal quantum computer~\cite{qgate1,qgate2,qgate3,qgate4}. By contrast, in a one-way quantum computer~\cite{oneway}, all entanglement resources for the quantum computation are provided initially in the form of special entangled states with a large number of qubits (so-called cluster states, and more generally, graph states)~\cite{cluster1,cluster2,cluster3}. The required teleportation of logical qubits and implementation of quantum gates can be simulated by one-qubit gates and one-qubit measurements on the cluster state in a polynomially efficient way. Another notable example is the zero-eigenvalue eigenstate (dark state) undergoing a one-qubit holonomic phase gate. Usually an entangled pure state of qubits will acquire a Berry's phase after the parameters of the Hamiltonian are tuned so that the state evolves along an adiabatic loop. Hence, it is desirable to have a robust protocol to create cluster or graph states for quantum information processing when investigating inverse engineering of an open quantum system.

Physically, atomic and molecular systems can experience a collective decay into a common environment, such as atom-like emitters and the nanophotonic waveguides~\cite{coll1,coll2}. Such systems, as we consider here, undergo collective motion from being acted upon by an environment that acts similarly on each individual qubit. One of the well-known properties of an environment that acts collectively is that the system can be completely decoupled from the environment when the system is in a DFS. Extra manipulation is sometimes needed to suppress the leakage of the system out of the DFS if the symmetry is not exact. In this case an optimal encoding scheme can be used for an approximate symmetry~\cite{Wang1,Wang2}.

In this work we investigate the possibility of using Lindblad dissipators~\cite{LM1,LM2} to create cluster/graph states on demand in a robust, deterministic and polynomially efficient way. The dissipative processes involving multiple interactions could exist in a state-of-the-art linear ion-trap quantum computing architecture. In particular, it has been shown that quantum information can be encoded in the set of steady states by the effective dissipators in master equation~\cite{Breuer,LM1,LM2}. These states are automatically free of decoherence since they are immune to the dissipators and our method does not rely on the symmetric coupling condition for conventional DFS. More significantly, working with dissipators opens an exotic branch of inverse engineering control in open quantum system which is robust against the influence of the quantum noise on the resource state itself due to the system-environment interaction. The stronger the coupling strength of the designed dissipators, the quicker the target state or subspace of the system is achieved.

\noindent \textbf{Results}\\
\textbf{Creating an open quantum system state.} In the field of open quantum system dynamics, the most general Markovian master equation can be written in a Lindblad form. Maintaining the translational invariance and positivity, the Lindblad master equation is an important and reliable tool for the treatment of irreversible and non-unitary processes of open-system states, covering the dissipation and pure-decoherence processes in the quantum measurement process. Even beyond the Markovian regime, it is practically used to find the steady state or subspace in the long-time limit for systems under stationary non-Markovian noises. Consider a system coupled to a multi-mode bosonic bath or field via the system operator $L$. Suppose the environment is at {\em zero temperature}, in the rotating frame we can arrive at a general master equation of the Lindblad form (see the derivation in \textbf{Method}),
\begin{equation}\label{Lindblad}
\pa_t\rho_S=\mathcal{L}\rho_S=
\ga\left[L\rho_SL^\da-\frac{1}{2}\left\{L^\da L, \rho_S\right\} \right].
\end{equation}
The Stark effect is ignored, which is not relevant to the non-unitary evolution of the system. Here $\ga=2\pi\int_0^\infty d\om J(\om)$ is the decay rate, where $J(\om)$ is the spectral function of the bath. The spectral function is obtained by the Fourier transformation of the two-point correlation function $f(t,s)=\sum_j|g_j|^2e^{-i\om_j(t-s)}$, where $g_j$ ($g_j^*$) is the coupling strength between system and the $j$-th mode with eigenfrequency $\om_j$. For a structured environment, $\ga$ could be time-dependent, but would become asymptotically time-independent for times longer than the timescale of correlation function. Equation~(\ref{Lindblad}) shows that the dark states of the system-environment interaction, i.e., those that satisfy $L|\Phi\ra=0$, will survive after the decoherence process determined by the dissipator $\mathcal{L}$. On the other hand, the dissipator serves as a filter to remove those states that are not in the steady-state subspace.

Extended to a more general situation with multiple environmental interactions, the microscopic Hamiltonian (see \textbf{Method}) becomes $H_I(t)=\sum_j[L_jB^\da_j(t)+h.c.]$. The right hand side of Lindblad equation~(\ref{Lindblad}) is thus generalized into a summation over dissipators with different $L_j$'s. Consider an $N$-dimensional system $\rho_S=\sum_{mn}\rho_{mn}|m\ra\la n|$, $\la m|n\ra=\de_{mn}$, which is under the irreversible dynamics determined by $L_{j>k}=|\varphi_j\ra\la j|$ with $|\varphi_j\ra=\sum_{p=1}^ka_{jp}|p\ra$ (it is not necessarily a normalized vector, so $a_{jp}$'s can be arbitrary). Evidently, we have
\begin{equation}\label{intui}
\mathcal{L}_j\rho_S=L_j\rho_SL_j^\da-\frac{1}{2}\left(L_j^\da L_j\rho_S+\rho_SL_j^\da L_j\right)
=\rho_{jj}|\varphi_j\ra\la\varphi_j|-\frac{||\varphi_j\ra|^2}{2}\sum_{n=1}^N
\left(\rho_{jn}|j\ra\la n|+\rho_{nj}|n\ra\la j|\right).
\end{equation}
Therefore the requirement of the long-time limit $\mathcal{L}\rho_S=\sum_{j=k+1}^N\mathcal{L}_j\rho_S=0$ will give rise to the vanishing of both population and coherence terms outside of the chosen $k\times k$ subspace of interest. When $k=1$, the system is eventually engineered to a target pure state, which is immune to the external disturbance and insensitive to the initial state.

Notably the effect of dissipator is invariant under simultaneous unitary transformation of the operators and states: $\ti{\mathcal{L}}\rho_S=U\mathcal{L}\rho_SU^\da$ with $\ti{L}=ULU^\da$ and $\ti{\rho}_S=U\rho_SU^\da$. This property aids in the experimental implementation. For instance, let us consider an open quantum two-level system with the target state an eigenstate $|\varphi\ra=|+\ra\equiv(|0\ra+|1\ra)/\sqrt{2}$ of the spin flip operator $X$ ($X|\pm\ra=\pm|\pm\ra$), where $X,Y,Z$ are Pauli matrices along the directions $x, y, z$, respectively. From the protocol in Eq.~(\ref{intui}), one can see that the Hamiltonian and Lindblad operator could be given by $H=\om X$ and $L=Z-iY$, respectively. After a sufficiently long time, this dissipator creates the superposed state in a single two-level system (one-dimensional DFS), $|+\ra$. In fact, if this is observed in the rotating frame with respect to the unitary transformation
\begin{equation*}
U=\sqrt{\frac{1}{2}}\left(\begin{array}{cc}1 & -1 \\ 1 & 1 \end{array}\right),
\end{equation*}
then this describes the well-known dissipative process for the single qubit system with $H\propto Z$ and $L\propto\si_-=(X-iY)/2$. Under the condition that the environmental degrees of freedom have a much faster relaxation timescale, the steady state of the system is governed by Eq.~(\ref{Lindblad}) and the preparation time would be sped up by strengthening the coupling between the system and environment, as allowed by the quantum speed limits~\cite{QSLO1,QSLO2}.

\textbf{Creating cluster state via dissipators.} Cluster or graph states are special class of states that are useful in quantum error-correcting codes, entanglement measurement and purification, and for characterization of computational resources in measurement based quantum computing models. The one-dimensional graph state consisting of $n$ qubits can be expressed as
\begin{equation}\label{graph}
|\varphi_n\ra=\frac{1}{2^{n/2}}\bigotimes_{q=1}^n
(|0\ra_qZ_{q+1}+|1\ra_q),
\end{equation}
with the convention $Z_{n+1}\equiv1$. For the two-qubit case, this may be written, up to local unitary transformations, as
\begin{equation}\label{twoq}
|\varphi_2\ra=\frac{1}{\sqrt{2}}(|0\ra_1|0\ra_2+|1\ra_1|1\ra_2).
\end{equation}
In quantum computing, a graph state can be represented by a qubit-network of $m$ nodes. Each node of this network (graph) is associated with a qubit prepared in the state of $|+\ra$ and each edge between two qubits $Q_1$ and $Q_2$ is acted on by a controlled phase gate, i.e., $|G_n\ra\equiv U|+\ra^{\otimes n}$, where $U=\prod_{Q_1,Q_2}U^{12}$ with $Q_1,Q_2\in\{1,2,3,\cdots,n\}$ and $U^{12}={\rm diag}([1,1,1,-1])$. For the two-qubit case, $|G_2\ra=|\varphi_2\ra$ up to a local unitary transformation. Similarly, one may obtain for $n=3,4$, $|\varphi_3\ra=(|000\ra+|111\ra)/\sqrt{2}$ and $|\varphi_4\ra=(|0000\ra+|0011\ra+|1100\ra-|1111\ra)/2$, respectively. The cluster states persist after being acted on by local measurements and constituents remain maximally connected.

To prepare a two-qubit system in the state $|\varphi_2\ra$ as in Eq.~(\ref{twoq}), we can employ three Lindblad operators $L_j=|\varphi_2\ra\la\phi_j|$, provided $\la\varphi_2|\phi_j\ra=0$ and $\la\phi_j|\phi_k\ra=\de_{jk}$, with $j,k=1,2,3$. For example, a set of possible solutions for the operators are found to be
\begin{eqnarray} \non
L_1&=&i(X_1Y_2+Y_1X_2)-(Z_1+Z_2), \\  \non L_2&=&i(Z_1Y_2+Y_1Z_2)+(X_1+X_2), \\ \label{Lfor2} L_3&=&(Z_1X_2-X_1Z_2)-i(Y_1-Y_2),
\end{eqnarray}
respectively, where the subscripts of $X,Y,Z$ imply the indexes of qubits. At the same line, it is intuitively clear that for creating a general cluster state with $n$ qubits, the operators $L_j$, $j=1,2,\cdots,2^n-1$, can be constructed as
\begin{equation}\label{Lj}
L_j\equiv\sum_\al W^{(j\al)}=\sum_\al W_1^{(j\al)} W_2^{(j\al)}\cdots W_n^{(j\al)},
\end{equation}
where $W_k\in\{X, Y, Z, I\}$, $I$ is the identity operator, and $\al$ is the index of the $n$-body interaction operators of order $O(n)$.

To be {\em polynomially efficient}, this method can be dramatically improved by utilizing a special Lindblad operator $L'=|\phi_0\ra(\sum_{\beta>0}a_\beta\la\phi_\beta|)$, where $|\phi_\beta\ra$'s constitute a complete set of eigenstates for the $n$-qubit system and $a_\beta$'s are non-vanishing constants. It is always possible to find a proper unitary transformation to get the cluster state $|\varphi_n\ra$ from a pure state connected to $|\phi_0\ra$. In another words, the protocol/algorithm based on Eq.~(\ref{Lj}) with many dissipators can be reformulated as one with a single dissipator. This special $L$ can be obtained as follows. The steady state determined by the vanishing dissipator in Eq.~(\ref{Lindblad}) can be always transformed into a diagonal form as $D=\mathcal{U}\rho_S\mathcal{U}^\da$, where $\mathcal{U}$ is a unitary matrix. Accordingly, $L$ is written as $L'=\mathcal{U}L\mathcal{U}^\da$. If $D=\sum_{\beta\geq0}P_\beta|\phi_\beta\ra\la\phi_\beta|$, then the Lindblad operator can be written as $L'=|\phi_0\ra(\sum_{\beta>0}a_\beta\la\phi_\beta|)$. Thus $\mathcal{L}'D=(\sum_{\beta>0}P_\beta|a_\beta|^2)|\phi_0\ra\la\phi_0|+
\sum_{\beta',\beta>0}P_{\beta'}a_\beta^*a_{\beta'}|\phi_\beta\ra\la\phi_{\beta'}|=0$, which yields $P_{\beta>0}=0$ for arbitrarily chosen coefficients $a_\beta$ so that the steady state is a pure state $|\phi_0\ra$. One should keep in mind that any pure state satisfying $L|\Phi\ra=0$ is a dark state of the dissipator of Eq.~(\ref{Lindblad}), and is therefore decoupled from the collective dissipation of the $n$-qubit system. However, the above specified $L'$ is not the unique solution of this type. Now returning to the previous frame, the Lindblad dissipator $L=\mathcal{U}^\da L'\mathcal{U}$ can be used to drive the system into a pure state $\mathcal{U}^\da|\phi_0\ra$ since the unitary transformation does not influence the purity of the steady state. Thus the conditions to realize Eq.~(\ref{Lj}) are relaxed to
\begin{equation}\label{LP}
L=\sum_\al W_1^{(\al)} W_2^{(\al)}\cdots W_n^{(\al)},
\end{equation}
{\em without} requiring an exponentially-increasing numbers of operations.

Physically, our protocol is available for a bosonic zero-temperature environment. One can use the above composite Lindblad dissipator to prepare the system in a pure state (not necessarily the ground state and could be the graph state $|\varphi_n\ra$) according to $|\varphi_n\ra=\mathcal{U}^\da|\phi_0\ra$. For the two-qubit cluster state in Eq.~(\ref{twoq}), the corresponding Lindblad operator is $L=\sum_{j=1}^3a_jL_j$, where $L_j$'s have been described in Eq.~(\ref{Lfor2}).

\begin{figure}[htbp]
\centering
\includegraphics[width=0.9\linewidth]{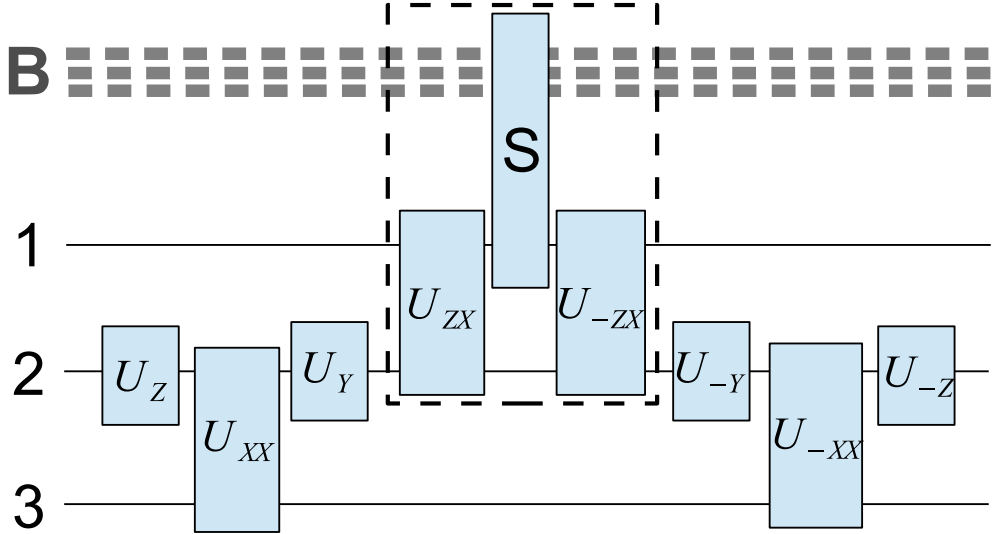}
\caption{(color online) Quantum circuit of realizing a special $3$-qubit interaction with the environment $B$ given by Eq.~(\ref{3qubits}). $T_{SB}=e^{i\theta Y_1B}$. $U_A=e^{i\frac{\pi}{4}A}$, where $A$ stands for Pauli operators $X_j$, $Y_j$, and $Z_j$ or their tensor products. }\label{circuit}
\end{figure}

\textbf{Creating many-body interactions.} The Lindblad master equation~(\ref{Lindblad}) is obtained from the partial trace over the total unitary transformation $U\rho_S\rho_BU^\da$. The full time-evolution operator $U=\mathcal{T}_{\leftarrow}\exp[-i\int_0^tdsH_I(s)]$ can be implemented using the Trotter formula by $U(\de_t)\approx\exp[-i(LB^\da+h.c.)\de_t]=\prod_{\al}
\exp[-i(W^{(\al)}B^\da+h.c.)\de_t]$. From Eq.~(\ref{Lindblad}), it is clear that the dissipator-creation control protocol must be based on the system operators in $\sum_{\al}W^{(\al)}B^\da$ that connect the system to the bath. The constituent $WB^\da$ (here for simplicity, the subscript of $W$ is omitted and $W$ is assumed to be unitary) involves a many-body interaction~\cite{Ari04,Oh} which is not naturally occurring. However, it can be constructed from the available interactions by a control protocol as follows. Let us start from the Hamiltonian describing the interaction of the first qubit and the environment $W_1B^\da$, which is always attainable. Then, supposing that the interaction between neighboring qubits $p, q$ contains a controllable term $W_pW_q$, such as $X_pX_q$ (in reality such control can be obtained as discussed below). For any set of operators satisfying $SU(2)$ commutation relations, such as $X,Y,Z$, we have the following useful formula:
\begin{equation*}
T_Z\circ e^{i\theta Y}\equiv e^{i(\pi/4)Z}e^{i\theta Y}e^{-i(\pi/4)Z}=e^{i\theta X},
\end{equation*}
where the operators could also be permuted cyclically, e.g., $Z\rightarrow Y\rightarrow X\rightarrow Z$. Using this property, we can efficiently establish the many-body interactions in an open system by alternately switching relevant interactions, e.g.,
\begin{equation}\label{3qubits}
T_{Z_2}\circ\{T_{X_2X_3}\circ[T_{Y_2}\circ(T_{Z_1X_2}\circ e^{i\theta Y_1B})]\},
\end{equation}
which can, in turn, be used to recursively generate $e^{i\theta X_1X_2X_3B}$. The process provided by Eq.~(\ref{3qubits}) is demonstrated by a circuit in Fig.~\ref{circuit} for the coupling term of $3$-qubits to the environment simultaneously, in which the dashed frame distinguishes the circuit generating $e^{i\theta X_1X_2B}$ for $2$-qubits coupled to the environment. A similar process can be used even for more long-range interactions of a many-qubit system coupled to a bosonic environment. Notably as shown in Fig.~\ref{circuit}, we need merely up to 2-qubit gates.

The two-body and even many-body interaction terms (quantum gates) have already been simulated in the ion-trap systems. For example, the time-evolution operator $U_{XX}=\exp(i\theta X_pX_q)$ can be implemented by two M{\o}lmer$-$S{\o}rensen gates~\cite{Simu1,Simu2} applied to the two system qubits (denoted as $p$ and $q$) and an ancilla qubit (denoted as $0$) initially prepared in $|0\ra$ along with a single-qubit rotation on the ancilla qubit, $U_{\rm MS}(-\pi/2,0)\exp(-i\theta Z_0)U_{\rm MS}(\pi/2,0)$, where $U_{\rm MS}(\mu,\nu)=\exp[-i\mu(\cos\nu S_x+\sin\nu S_y)^2/4]$, $S_x=X_0+X_p+X_q$ and $S_y=Y_0+Y_p+Y_q$. Using properties of the group $SU(2)$, arbitrary two-body interactions can be simulated with the help of local unitary transformations by quantum gates~\cite{LW}.

\noindent\textbf{Discussion} \\
Dissipators in the Markovian master equation or non-Markovian master equation presented in the Lindblad form are usually regarded as decoherence or non-unitary evolution of open quantum systems which are detrimental. In this work, we have shown how to efficiently guide an open quantum system into one of the zero eigenvalue eigenstates (dark states) by using certain dissipators acting as generators of the desired state or subspace.

In particular, we have presented an explicit control protocol to create cluster states, or graph states, suitable for one-way quantum computing on an $n$-qubit system. Using circuit diagrams with up to the two-qubit quantum gates, the relevant Lindblad operators as well as the full evolution operator can be decomposed into $O(n)$ elements of many-body interactions between qubits and the bath. Our protocol can be realized in a {\em polynomially} efficient way and could also be implemented using available quantum gates in ion-trap systems. Possible extensions of this work include a protocol to generate subsystem codes and to optimize the operations to improve the inverse engineering efficiency of the desired states.

\noindent\textbf{Method}\\
The total Hamiltonian describing both system and multi-mode environment in the rotating frame can be generally written as
\begin{equation}\label{HI}
H_I(t)=LB^\da(t)+L^\da B(t),
\end{equation}
where $L$ and $B(t)=\sum_jg_ja_je^{-i\om_jt}$ are the Lindblad operator and the bath operator, respectively. $a_j$ is the annihilation operator for the $j$-th environmental mode with eigenfrequency $\om_j$ and $g_j$ is its coupling strength with the system. The corresponding Born-Markov master equation reads
\begin{equation}\label{Born}
\pa_t\rho_S(t)=-\int_0^{\infty}ds{\rm Tr}_B\left\{[H_I(t), [H_I(s), \rho_S(t)\rho_B]]\right\},
\end{equation}
which is based on a popular assumption of no initial correlations between the system and environment. Tracing over the environmental degrees (${\rm Tr}_B$) plays the role of an average measurement over the environment. A straightforward derivation from Eq.~(\ref{Born}) yields
\begin{equation*}
\pa_t\rho_S=-[F(t)(L^\da L\rho_S-L\rho_SL^\da)+G(t)(\rho_SLL^\da-L^\da\rho_SL)+h.c.],
\end{equation*}
where $F(t)=\int_0^{\infty}ds{\rm Tr}_B[B(t)B^\da(s)\rho_B]$ and $G(t)=\int_0^{\infty}ds{\rm Tr}_B[B^\da(t)B(s)\rho_B]$. Suppose the environment is at {\em zero temperature}, then $F(t)=\sum_j|g_j|^2\int_0^tdse^{-i\om_j(t-s)}$ and $G(t)=0$. Ignoring the Stark effect, which is not relevant to the non-unitary evolution of the system, we arrive at a master equation of the Lindblad form,
\begin{equation}\label{Lindblad1}
\pa_t\rho_S=\mathcal{L}\rho_S=
\ga\left[L\rho_SL^\da-\frac{1}{2}\left\{L^\da L, \rho_S\right\} \right],
\end{equation}
where $\ga=2\pi\int_0^\infty d\om J(\om)$ is the decay rate. Here $J(\om)$ is the spectral function of the bath coupled to the system via the system operator $L$, which is obtained by the Fourier transformation of the two-point correlation function $f(t,s)=\sum_j|g_j|^2e^{-i\om_j(t-s)}$.

The universality of Eq.~(\ref{Lindblad1}) or Eq.~(\ref{Lindblad}) can alternatively be shown using the quantum-state-diffusion (QSD) equation~\cite{qsd1,qsd2} which is a dynamical equation for the stochastic wavefunction of the system subject to the influence from the environment. Starting from Eq.~(\ref{HI}) and assuming a zero-temperature environment, the QSD equation reads,
\begin{equation}\label{qsd}
\pa_t\psi_t(z^*)=\left[Lz_t^*-L^\da\int_0^tdsf(t,s)O(t,s,z^*)\right]\psi_t(z^*).
\end{equation}
The stochastic process $z_t^*\equiv-i\sum_jg_j^*z_j^*e^{i\om_jt}$ is the result of the aforementioned partial trace over the environment, where $z_j^*$ is a Gaussian random number indicating a random coherent state of the $j$-th environmental mode. The system operator $O(t,s,z^*)$ embodies the system-environment interaction. The density matrix of the system is obtained by ensemble average $\rho_S=M[|\psi_t\ra\la\psi_t|]$. Using a Markov approximation, $O(t\rightarrow s,s,z^*)\rightarrow L$, which corresponds to the environmental correlation function with $f(t,s)=\ga\de(t-s)$. In this limit, the Novikov theorem will ensure that Eq.~(\ref{qsd}) can be used to represent the same open dynamics as indicated by Eq.~(\ref{Lindblad}).

\textbf{Acknowledgements}\\
We acknowledge grant support from the Basque Government (grant IT472-10), the Spanish MICINN (No. FIS2012-36673-C03-03), the National Science Foundation of China No. 11575071, and UMass Boston (P20150000029279).

\textbf{Author contributions}\\
J.J. performed analyzed results and prepared figures. L.-A.W. contributed to the conception and development of the research problem. All authors (J.J., M.S.B. and L.-A.W.) discussed the results and physical implications, and wrote the manuscript.

\textbf{Additional Information} \\
Competing financial interests: The authors declare no competing financial interests.

\end{document}